\def\bar{\overline}

\def\k{\kappa}
\def\e{\epsilon}

\def\bar{\overline}
\def\l{\lambda}
\def\G{{\rm GeV}}

\def\eV{{\rm eV}}
\documentstyle [12 pt]{article}
\begin{document}
\baselineskip=22 pt
\setcounter{page}{1}
\thispagestyle{empty}
\topskip 0.5  cm
\centerline{\LARGE \bf    Seesaw Enhancement  of Bi-large Mixing}
\vskip 0.5 cm
\centerline{\LARGE \bf  in Two-Zero Textures}
\vskip 1.5 cm
\centerline{{\large \bf Mizue Honda}
 \renewcommand{\thefootnote}{\fnsymbol{footnote}}
\footnote[1]{E-mail address:  mizue@muse.sc.niigata-u.ac.jp},
\qquad{\large \bf Satoru Kaneko}
\renewcommand{\thefootnote}{\fnsymbol{footnote}}
\footnote[2]{E-mail address: kaneko@muse.sc.niigata-u.ac.jp}}
\vskip 0.3 cm
\centerline{\large\bf and}
\vskip 0.3 cm
\centerline{{\large \bf Morimitsu Tanimoto}
\renewcommand{\thefootnote}{\fnsymbol{footnote}}
\footnote[3]{E-mail address: tanimoto@muse.sc.niigata-u.ac.jp}
 }
\vskip 0.8 cm
 \centerline{ \it{Department of Physics, Niigata University, 
 Ikarashi 2-8050, 950-2181 Niigata, JAPAN}}
\vskip 1.5 cm
\centerline{\bf ABSTRACT}\par
\vskip 0.8 cm
The seesaw enhancement of the bi-large mixings are discussed for the two-zero
textures of the neutrino mass matrix. There are no large mixings in both 
Dirac neutrino mass matrix $m_D$ and right-handed Majorana neutrino mass 
matrix $M_R$, however, the bi-large mixing is realized via the seesaw 
mechanism. We present twelve sets of $m_D$ and $M_R$ for the seesaw 
enhancement and discuss the related phenomena, the $\mu\rightarrow e+\gamma$
process and the leptogenesis. The decay rate of $\mu\rightarrow e+\gamma$ is 
enough suppressed due to zeros in the Dirac neutrino mass matrix. Six  $m_D$ 
lead to  the lepton asymmetry, which can explain the baryon number in the 
universe.  Other six  $m_D$ are the real matrices, which give no CP 
asymmetry. Modified Dirac neutrino mass matrices are also discussed.
\newpage
\topskip 0. cm
The texture with zeros of the neutrino mass matrix have been discussed 
\cite{FuNi,AK,Kang,Chen}
 to explain  neutrino masses and mixings \cite{MNS}, which have been presented
by the recent neutrino experiments  \cite{SKam,SKamsolar,SNO,KamLAND}.
It was found that the two-zero textures  are consistent with 
the experimental data in the basis of the diagonal charged lepton mass 
matrix \cite{Fram}. Consequently, the neutrino mass matrix does not display 
the hierarchical  structure as seen in the quark mass matrix
 \cite{Xing1,Xing2,Barbieri,Chen2,Obara,Ibara,HKT}.

Since the two-zero textures of ref.\cite{Fram} are given for the light effective neutrino mass matrix $M_\nu$, one needs to find the seesaw realization 
\cite{Seesaw} of these textures from the standpoint of the model building.
We have examined the seesaw realization of the neutrino mass matrix
with two zeros \cite{our}.  
Without fine tunings between parameters  of the Dirac neutrino mass matrix 
$m_D$ and the right-handed Majorana neutrino one $M_R$,
we obtained several textures  of $m_D$ for the  fixed  $M_R$ \cite{our}.
Among them, there are textures  of $m_D$ and $M_R$ which have 
 hierarchical masses without large mixings.
These   present the seesaw enhancement of  mixings,
because there is no large mixings in  $m_D$ and $M_R$,
but the bi-large mixing is  realized via the seesaw mechanism.
The seesaw enhancement  are important in the standpoint of the 
quark-lepton unification, in which  quark masses are hierarchical and 
quark mixings are very small
\footnote{Although phenomenological analyses of the two-zero textures were 
given in the diagonal basis of the charged lepton, some authors 
\cite{Obara,Ibara,HKT} have also studied the  two-zero textures of neutrinos 
in the basis of charged lepton mass matrix with small off-diagonal components.}.

The general discussions of the seesaw enhancement were given 
in the case of two flavors \cite{enhancement1,enhancement2}. 
Specific cases  were discussed in the case of three flavors 
\cite{enhancement3,enhancement4} because it is very difficult to get general 
conditions for the seesaw enhancement of the bi-large mixing.

However, the two-zero texture of the neutrino mass matrix $M_\nu$ are helpful 
 to study the seesaw enhancement of  the bi-large mixing. 
In this paper, 
we present sets of  $m_D$ and $M_R$ to give the seesaw enhancement 
in the two-zero textures of $M_\nu$ and discuss the related phenomena,
the $\mu\rightarrow e+\gamma$ process and the leptogenesis \cite{FuYa}.
 
There are fifteen two-zero textures for the  neutrino 
mass matrix  $M_\nu$, which have five independent parameters. 
 Among these textures, seven acceptable textures with two independent 
zeros were found for the  neutrino mass matrix \cite{Fram}, 
and they have been studied in detail \cite{Xing2,Barbieri,Ibara,HKT}.
Especially, the textures $A_1$ and $A_2$ of ref.\cite{Fram}, which  
correspond to the hierarchical neutrino mass spectrum, are strongly favored 
 by the recent phenomenological analyses \cite{Xing1,Xing2,HKT}.
Therefore, the two textures are taken in order to discuss the seesaw 
enhancement.

Putting data of neutrino masses and mixings  \cite{Lisi},
\begin{eqnarray}
 0.35 \le  \tan^2 \theta_{\rm sun} \le 0.54  \ , \ \ 
&&6.1\times 10^{-5} \le  \Delta m^2_{\rm sun} \le  
8.3\times 10^{-5}~\rm{eV}^2,  
 \quad 90\% C.L. \ ,\nonumber \\ 
0.90 \le \sin^2 2\theta_{\rm atm}  \ , \ \ 
&& 1.3\times 10^{-3} \le  \Delta m^2_{\rm atm} \le 3.0 \times 10^{-3}~
 \rm{eV}^2 ,  \quad 90\% C.L. \ ,
\label{Data}
\end{eqnarray}
the relative magnitude of each entry of the neutrino mass matrix is roughly
given for  the textures $A_1$ and $A_2$  as follows: 
\begin{eqnarray}
M_\nu \ \simeq \ \  m_0 \left ( \matrix{
{ 0} & {0} & \lambda \cr {0} & 1 & 1 \cr \lambda  & 1 & 1 \cr} \right )
\  \ \ {\rm for}\  A_1 \ , \qquad
\qquad 
m_0 \left (\matrix{ 0 & \lambda & 0\cr\lambda  & 1 & 1 \cr 0 & 1 & 1 \cr} 
\right )
\  \ \ {\rm for} \ A_2 \ ,
\label{A12}
\end{eqnarray}
\noindent
where  $m_0$ denotes  a constant mass  and $\lambda\simeq 0.2$.  
These matrix is given in terms of the Dirac neutrino mass matrix 
${m_D}$ and the right-handed Majorana neutrino mass matrix ${ M_R}$ 
by the  the seesaw mechanism as
\begin{equation}
 M_\nu=  m_D\   M_R^{\rm -1}\  m_D^{\rm T} \ .
\end{equation} 
\noindent 
Zeros in $m_D$  and $M_R$ provide  zeros
in the neutrino mass matrix $M_\nu$ of  eq.(\ref{A12})
as far as we exclude the possibility that  zeros are originated
 from accidental cancellations among matrix elements. 
  In other words, we take a standpoint that  the two-zero texture should come 
from zeros of the Dirac neutrino mass matrix and the right-handed Majorana
  mass matrix.
Possible  textures  of  $m_D$ and $M_R$  were given in ref. \cite{our}.
Among them, we select the set of  $m_D$ and $M_R$, which reproduce the
seesaw enhancement of the bi-large mixing.

 Let us fix the right-handed Majorana neutrino mass matrix without 
large mixings.
We take  simple right-handed Majorana neutrino mass matrix with 
only three independent parameters. Then, there are  $_6{\rm C}_3=20$ textures.
Among them, six textures are excluded because they have a  zero eigenvalue, 
which corresponds to a massless right-handed Majorana neutrino. 
Other two textures are also excluded because the two-zero textures $A_1$ and 
$A_2$ cannot be reproduced without accidental cancellations.  
One of the two textures is the diagonal matrix, 
and another one is the matrix  with three zeros in the  diagonal elements.

We show  twelve  real mass matrices  with three independent parameters
\footnote{The classification of $M_R$, types $a_i$,  $b_i$, $c_i$ 
($i=1,2,3,4$), follows from  ref.\cite{our}.}
 with mass eigenvalues
$|M_1|= \l^m M_3$ and $|M_2|= \l^n M_3$, where  $M_3$ is the mass of the
third generation,  and $m$ and $n$ are integers with $m > n > 1$: 
\begin{eqnarray}
{ a_i\  type} \quad 
M_R\simeq &&  M_3\left ( \matrix{
-1&0 &\lambda^{\frac{m}{2}}\cr 0& \l^n & 0\cr \lambda^{\frac{m}{2}}& 0& 0\cr} 
 \right )_{a_1}  \ ,  \qquad 
  M_3\left( \matrix{
0&-\l^{\frac{n}{2}} &\lambda^{\frac{m+n}{2}}\cr  -\l^{\frac{n}{2}}& 1& 0\cr
 \lambda^{\frac{m+n}{2}}& 0& 0\cr} 
 \right )_{a_2} \ , \nonumber \\
&& 
 M_3\left ( \matrix{
0&0 &\lambda^{\frac{m}{2}}\cr 0& \l^n & 0\cr \lambda^{\frac{m}{2}}& 0& -1\cr} 
 \right )_{a_3} \ ,  \qquad 
  M_3\left( \matrix{
0&0 &\lambda^{\frac{m+n}{2}}\cr  0& 1& -\l^{\frac{n}{2}}\cr
 \lambda^{\frac{m+n}{2}}& -\l^{\frac{n}{2}}& 0\cr} 
 \right )_{a_4}  \ ,
\label{MRa1}
\end{eqnarray}
\begin{eqnarray}
{ b_i\  type} \quad 
M_R\simeq &&  M_3\left ( \matrix{
-\l^n&\lambda^{\frac{m+n}{2}}&0\cr \lambda^{\frac{m+n}{2}}&0 & 0\cr
  0& 0 &1\cr} 
 \right )_{b_1}  \ ,  \qquad 
  M_3\left( \matrix{
0&\l^{\frac{m+n}{2}} &-\lambda^{\frac{n}{2}}\cr  \l^{\frac{m+n}{2}}& 0& 0\cr
 -\lambda^{\frac{n}{2}}& 0& 1\cr} 
 \right )_{b_2} \ , \nonumber \\
&& 
 M_3\left ( \matrix{
0&\lambda^{\frac{m+n}{2}}&0\cr \lambda^{\frac{m+n}{2}}& -\l^n & 0\cr 
0& 0& 1\cr} \right )_{b_3} \ ,  \qquad 
 M_3 \left( \matrix{
0&\lambda^{\frac{m+n}{2}}&0\cr \lambda^{\frac{m+n}{2}}& 0&-\l^{\frac{n}{2}}\cr
0& -\lambda^{\frac{n}{2}}& 1\cr} 
 \right )_{b_4} \ ,
\label{MRb1}
\end{eqnarray}
\begin{eqnarray}
{c_i\  type} \quad 
M_R\simeq &&  M_3\left ( \matrix{
\l^m&0 &0\cr 0&-1&\lambda^{\frac{n}{2}} \cr
  0& \lambda^{\frac{n}{2}} &0\cr} 
 \right )_{c_1} \ ,  \qquad 
  M_3\left( \matrix{
1&-\lambda^{\frac{n}{2}}&0\cr  -\lambda^{\frac{n}{2}}&0&\l^{\frac{m+n}{2}}\cr
 0& \l^{\frac{m+n}{2}}& 0\cr} 
 \right )_{c_2}  \ , \nonumber \\
&& 
 M_3\left ( \matrix{
\l^m&0 &0\cr 0&0&\lambda^{\frac{n}{2}} \cr
  0& \lambda^{\frac{n}{2}} &-1\cr} 
 \right )_{c_3} \ ,  \qquad 
  M_3\left( \matrix{
1&0&-\lambda^{\frac{n}{2}}\cr 0&0&\l^{\frac{m+n}{2}}\cr
 -\lambda^{\frac{n}{2}}& \l^{\frac{m+n}{2}}& 0\cr} 
 \right )_{c_4}  \ ,
\label{MRc1}
\end{eqnarray}
\noindent
where there are no large mixings  in twelve  matrices
since  the mass eigenvalues are supposed to be hierarchical.
The minus signs in the matrix elements are taken  to reproduce signs
 in the texture $A_1$ and $A_2$ of eq.(\ref{A12}).

 There are several  Dirac neutrino mass matrices to give 
 the textures  $A_1$ and $A_2$ in eq.(\ref{A12}) \cite{our}.
 We show    Dirac neutrino mass matrices $(m_D)_{a_i}$,  $(m_D)_{b_i}$, 
$(m_D)_{c_i}$ ($i=1\sim 4$) with maximal number of  zeros,  
which have no large mixings, to  give the texture $A_2$
\footnote{
For the  texture $A_1$, we easily obtain the Dirac neutrino mass matrices
 by exchanging the second  and third rows.}.
For each matrix of $(M_R)_{a_i}$, $(M_R)_{b_i}$, $(M_R)_{c_i}$ those are given
as follows:
\begin{eqnarray}
{ a_i\  type} \quad 
m_D\simeq &&  m_{D0}\left ( \matrix{
\l&0 &0\cr 0&0&\lambda^{\frac{m}{2}}\cr 1& \l^{\frac{n}{2}}& 0\cr} 
 \right )_{a_1} \ ,  \qquad 
  m_{D0}\left( \matrix{
\l^{\frac{n}{2}+1} &0&0\cr 0&0&\lambda^{\frac{m}{2}}\cr 0& 1& 0\cr} 
 \right )_{a_2}   \ , \nonumber \\
\label{mDa1}\\
&& 
 m_{D0}\left ( \matrix{
0&0 &\lambda\cr \lambda^{\frac{m}{2}}& 0&0\cr0&\lambda^{\frac{n}{2}} & 1\cr} 
 \right )_{a_3} \ ,  \qquad 
  m_{D0}\left( \matrix{
0&0 &\lambda^{\frac{n}{2}+1}\cr  \lambda^{\frac{m}{2}}& 0& 0\cr 0& 1& 0\cr} 
 \right )_{a_4}  \ ,  \nonumber
\end{eqnarray}

\begin{eqnarray}
{ b_i\  type} \quad 
m_D\simeq &&  m_{D0}\left ( \matrix{
\l^{\frac{n}{2}+1}&0 &0\cr   0&\lambda^{\frac{m}{2}}&0\cr
\l^{\frac{n}{2}}& 0&1\cr} 
 \right )_{b_1}  \ ,  \qquad 
  m_{D0}\left ( \matrix{
\l^{\frac{n}{2}+1}&0 &0\cr  0&\lambda^{\frac{m}{2}}&0\cr0& 0&1\cr} 
 \right )_{b_2} \ , \nonumber \\  
\label{mDb1}\\
&& 
 m_{D0}\left ( \matrix{
0&\l^{\frac{n}{2}+1}&0\cr \lambda^{\frac{m}{2}}& 0&0\cr
0&\lambda^{\frac{n}{2}} & 1\cr} 
 \right )_{b_3}  \ ,  \qquad 
  m_{D0}\left( \matrix{
0&\l^{\frac{n}{2}+1}&0\cr  \lambda^{\frac{m}{2}}& 0&0\cr 0&0 & 1\cr} 
 \right )_{b_4}  \ ,  \nonumber
\end{eqnarray}

\begin{eqnarray}
{ c_i\  type} \quad 
m_D\simeq &&  m_{D0}\left ( \matrix{
0&\l &0\cr   0&0&\lambda^{\frac{n}{2}}\cr \l^{\frac{m}{2}}& 1&0\cr} 
 \right )_{c_1}  \ ,  \qquad 
  m_{D0}\left ( \matrix{
0&\l^{\frac{n}{2}+1}&0 \cr   0&0&\lambda^{\frac{m}{2}}\cr1& 0&0\cr} 
 \right )_{c_2} \ , \nonumber \\
\label{mDc1}\\
&& 
 m_{D0}\left ( \matrix{
0&0&\l \cr    0&\lambda^{\frac{n}{2}}&0\cr \l^{\frac{m}{2}}& 0&1\cr} 
 \right )_{c_3}  \ ,  \qquad 
  m_{D0}\left( \matrix{
0& 0&\l^{\frac{n}{2}+1}\cr   0&\lambda^{\frac{m}{2}}&0\cr 1& 0&0\cr} 
 \right )_{c_4}  \ ,  \nonumber
\end{eqnarray}
\noindent
where   $m_{D0}$ denotes the magnitude  of the Dirac neutrino mass
and  complex  coefficients of order one  are omitted.
Although these matrices have no large mixing among three families,
the  neutrino mass matrix $M_\nu$ has the bi-large mixing through the seesaw
mechanism. These are so called  seesaw enhancement of the  bi-large mixing.

These Dirac matrices are ones with maximal number of  zeros.  
Without changing the mixings and the mass eigenvalues in the leading order, 
some  zeros can be replaced with small {\it non-zero} entries as follows:
\begin{eqnarray}
{ a_i\  type} \ \ 
m_D\simeq &&  m_{D0}\left ( \matrix{
\l&0 &0\cr \l^x & \l^y&\lambda^{\frac{m}{2}}\cr 1& \l^{\frac{n}{2}}& 0\cr} 
 \right )_{a_1}  ,  \  \matrix{ x>0\cr  y>\frac{n}{2}}\ , \quad 
  m_{D0}\left( \matrix{
\l^{\frac{n}{2}+1} &0&0\cr \l^x&\l^z&\lambda^{\frac{m}{2}}\cr \l^y& 1& 0\cr} 
 \right )_{a_2} , \ \matrix{ x>\frac{n}{2}\cr y>\frac{n}{2}\cr z>0}\nonumber\\
\label{mDa1x}\\
&& 
 m_{D0}\left ( \matrix{
0&0 &\lambda\cr \lambda^{\frac{m}{2}}& \l^x& \l^y
 \cr0&\lambda^{\frac{n}{2}} & 1\cr} 
 \right )_{a_3} ,  \  \matrix{ x>\frac{n}{2}\cr  y>0} \ , \quad 
  m_{D0}\left( \matrix{
0&0&\lambda^{\frac{n}{2}+1}\cr\lambda^{\frac{m}{2}}&\l^z&\l^x\cr 0&1&\l^y\cr} 
 \right )_{a_4} , \ \matrix{ x>\frac{n}{2}\cr y>\frac{n}{2}\cr z>0} \nonumber
\end{eqnarray}

\begin{eqnarray}
{ b_i\  type} \ \
m_D\simeq &&  m_{D0}\left ( \matrix{
\l^{\frac{n}{2}+1}&0 &0\cr   \l^x&\lambda^{\frac{m}{2}}&\l^y\cr
\l^{\frac{n}{2}}& 0&1\cr} 
 \right )_{b_1}  , \ \matrix{ x>\frac{n}{2}\cr  y>0}\ , \quad 
  m_{D0}\left ( \matrix{
\l^{\frac{n}{2}+1}&0 &0\cr  \l^x&\lambda^{\frac{m}{2}}&\l^z\cr \l^y& 0&1\cr} 
 \right )_{b_2} ,\  \matrix{ x>\frac{n}{2}\cr  y>\frac{n}{2}\cr z>0}\nonumber \\
 \label{mDb1x}\\
&& 
 m_{D0}\left ( \matrix{
0&\l^{\frac{n}{2}+1}&0\cr \lambda^{\frac{m}{2}}& \l^x&\l^y\cr
0&\lambda^{\frac{n}{2}} & 1\cr} 
 \right )_{b_3}   ,  \  \  \matrix{ x>\frac{n}{2}\cr  y>0}\ ,\quad 
  m_{D0}\left( \matrix{
0&\l^{\frac{n}{2}+1}&0\cr  \lambda^{\frac{m}{2}}& \l^x&\l^z\cr 0&\l^y & 1\cr} 
 \right )_{b_4} , \ \matrix{ x>\frac{n}{2}\cr y>\frac{n}{2}\cr z>0} \nonumber
\end{eqnarray}

\begin{eqnarray}
{ c_i\  type} \ \
m_D\simeq &&  m_{D0}\left ( \matrix{
0&\l &0\cr   \l^x&\l^y&\lambda^{\frac{n}{2}}\cr \l^{\frac{m}{2}}& 1&0\cr} 
 \right )_{c_1}   , \  \matrix{ x>\frac{m}{2}\cr  y>0}\ , \quad 
  m_{D0}\left ( \matrix{
0&\l^{\frac{n}{2}+1}&0 \cr   \l^x&\l^y&\lambda^{\frac{m}{2}}\cr1& \l^z&0\cr} 
 \right )_{c_2} ,\ \matrix{x>0\cr y>\frac{n}{2}\cr z>\frac{n}{2}}\nonumber \\
\label{mDc1x}\\
&& 
 m_{D0}\left ( \matrix{
0&0&\l \cr    \l^x&\lambda^{\frac{n}{2}}&\l^y\cr \l^{\frac{m}{2}}& 0&1\cr} 
 \right )_{c_3}   ,  \  \matrix{ x>\frac{m}{2}\cr  y>0} \ , \quad
  m_{D0}\left( \matrix{
0& 0&\l^{\frac{n}{2}+1}\cr   \l^x&\lambda^{\frac{m}{2}}&\l^y\cr 1& 0&\l^z\cr} 
 \right )_{c_4} , \ \matrix{x>0\cr y>\frac{n}{2}\cr z>\frac{n}{2}}\nonumber
\end{eqnarray}
\noindent
where $x$, $y$ and $z$ are positive integers.
These Dirac neutrino mass matrices are asymmetric ones.  However,
 only the $b_3$ and $b_4$ textures in eq.(\ref{mDb1x}) 
are adapted to the symmetric texture in the $SO(10)$-like GUT 
if $y$, $m$ and $n$ are relevantly chosen \cite{Obara,Chen2}.
For example, in  the  $b_3$  case, taking  $y=n/2$ and  $m=n+2$, 
the symmetric mass matrix is 
given, especially, putting $n=8$, we have the hierarchical mass 
matrix such like the up-quark  mass matrix.


Let us discuss these  obtained textures of $m_D$ in the
 $\mu\rightarrow e +\gamma$ decay and the leptogenesis.
It is well known that the Yukawa coupling of the neutrino contributes to
 the lepton flavor violation (LFV).  
Many authors have studied the LFV in the minimal supersymmetric standard 
model (MSSM) with right-handed neutrinos assuming the relevant neutrino 
mass matrix \cite{Borz,LFV1,LFV2,Sato,Casas,ourx}.
In the MSSM with soft breaking terms, 
there exist  lepton flavor violating terms such as 
off-diagonal elements of slepton mass matrices 
and trilinear couplings (A-term).
It is noticed that large neutrino Yukawa couplings and large lepton mixings 
 generate the large LFV in the left-handed slepton masses. For example, 
the decay rate of  $\mu\rightarrow e +\gamma$ can be approximated as follows:
\begin{eqnarray}
{\rm \Gamma} (\mu\rightarrow e +\gamma) \simeq 
\frac{e^2}{16 \pi} m^5_{\mu} F
 \left| \frac{(6+2 a_0^2)m_{S0}^2}{16 \pi^2}
({Y_\nu  Y_\nu^\dagger})_{21} \ln\frac{M_X}{M_R}  \right|^2\ ,
\label{rate}
\end{eqnarray}
\noindent where the neutrino Yukawa coupling matrix  $ Y_\nu$ 
 is given as  ${Y_\nu}={m_D}/v_2$ ($v_2$ is a VEV of Higgs)
at the right-handed mass scale $M_R$, and 
  $F$ is a function of masses and mixings for SUSY particles.  
In eq.(\ref{rate}),
 we assume the universal scalar mass $(m_{S0})$ for all scalars and
the universal A-term $(A_f=a_0 m_{S0} Y_f)$ at the GUT scale $M_X$. 
Therefore  the branching ratio $\mu \rightarrow e +\gamma$ depends 
considerably on the texture $m_D$ \cite{Sato,Casas,ourx}.

The magnitude of $({ m_D m_D^\dagger})_{21}$ is a key ingredient
to predict the branching ratio of the $\mu\rightarrow e+ \gamma$ process
\footnote{${ m_D m_D^\dagger}$ does not depend 
on the basis of the right-handed sector.}.
Many works have shown that this branching ratio is too large 
\cite{Sato,Casas}.
The conditions for  $({ m_D m_D^\dagger})_{21}$ were given
in ref. \cite{Ellis} as follow:
\begin{eqnarray}
 H_{21}  \leq
 10^{-2}\times \tan^{-\frac{1}{2}}\beta \left (\frac{m_{S0}}{100\G} \right )^2
 \left( \frac{Br(\mu\rightarrow e\gamma)}{1.2\times 10^{-11}}
\right )^{-\frac{1}{2}} \ , \nonumber\\
 H_{31}  H_{23} \leq
 10^{-1}\times \tan^{-\frac{1}{2}}\beta \left (\frac{m_{S0}}{100\G} \right )^2
 \left( \frac{Br(\mu\rightarrow e\gamma)}{1.2\times 10^{-11}}
\right )^{-\frac{1}{2}} \ ,
\label{cond}
\end{eqnarray}
 where
\begin{eqnarray}
H_{ij}= \sum_{k}{(m_D)_{ik}(m_D^\dagger)}_{kj}  \ln \frac{M_X}{M_{Rk}} \ .
\end{eqnarray}
These conditions give constraints for the magnitude of 
$({ m_D m_D^\dagger})_{ij}$  and $M_3$.
Zeros in  Dirac mass  matrices $m_D$ may lead to  
$({ m_D m_D^\dagger})_{ij}=0$  and then it suppress the
 $\mu\rightarrow e +\gamma$ decay.
Actually, all $m_D$ in eqs.(\ref{mDa1}), (\ref{mDb1}) and 
(\ref{mDc1}) give $({ m_D m_D^\dagger})_{21}= 0$  and
 $({ m_D m_D^\dagger})_{31}({ m_D m_D^\dagger})_{23}= 0$
\footnote{For the texture $A_1$ case, the some Dirac mass matrices
give non-zero $({ m_D m_D^\dagger})_{21}$, which leads to the constraint
 for  $M_3$}.
Even if non-zero terms $\l^x$, $\l^y$, $\l^z$ are taken 
as seen in eqs.(\ref{mDa1x}), (\ref{mDb1x}) and (\ref{mDc1x}),
$({ m_D m_D^\dagger})_{21}$  and
 $({ m_D m_D^\dagger})_{31}({ m_D m_D^\dagger})_{23}$ are suppressed
as far as $x,y,z \gg 1$.
Then, the branching ratio is safely predicted to be below the present 
 experimental upper bound $1.2\times 10^{-11}$ \cite{exp} due to zeros.

 Let us examine our textures in the leptogenesis 
 \cite{lept1,lept2,lept3}, which is based on the Fukugita-Yanagida
mechanism \cite{FuYa}.
 The CP violating phases in the Dirac neutrino mass matrix are key ingredients
for the leptogenesis while
the right-handed Majorana neutrino mass matrix are taken to be real
in eqs.(\ref{MRa1}), (\ref{MRb1}) and (\ref{MRc1}).
 Although the non-zero entries  in the Dirac neutrino mass matrix
are  complex, three phases are removed 
by the re-definition of the left-handed neutrino fields.
There is no freedom of re-definition for the right-handed ones
in the basis with  the real $ M_R$.
We should move to the  diagonal basis of the right-handed Majorana 
neutrino mass matrix in order to calculate the magnitude of the leptogenesis.
Then, the Dirac neutrino mass matrices $\bar m_D$ in the new basis
is given  as follows:
\begin{equation}
  { \bar m_D} =  { P_L \  m_D \ O_R} \ ,
\end{equation}
\noindent
where $ P_L$ is a diagonal phase matrix and $O_R$ is the 
orthogonal matrix which diagonalizes $M_R$ as  $ O_R^T M_R O_R$
in eqs.(\ref{MRa1}),  (\ref{MRb1}) and (\ref{MRc1}).
Since  the phase matrix $ P_L$ can remove one phase in each row of
 $m_D$, three phases disappear  in ${\bar m_D}$.

As a typical example, we show the case of the $b_3$ texture 
in eq.(\ref{MRb1}).  By taking three eigenvalues of $ M_R$  as follows 
\footnote{The minus sign of $M_2$ is necessary to reproduce 
$ M_{R}$  in eq.(\ref{MRb1}).  This minus sign is transfered to  
$ m_D$ by the right-handed diagonal phase matrix $diag(1,i,1)$.}:
\begin{equation}
  M_1=\l^m  M_3  \  , \qquad M_2= -\l^n  M_3  \ . 
\end{equation} 
 We obtain the orthogonal matrix $O_R$ as 
\begin{equation}
  {O_R} =  \left ( \matrix{ \cos{\theta} &  \sin{\theta}&  0\cr 
 - \sin{\theta} &  \cos{\theta} & 0 \cr  0 &  0  & 1 } \right )  \ ,
\qquad\quad  \tan^2 {\theta}= \l^{m-n} \ .
\end{equation}
 Then the  Dirac  mass matrices  $m_D$ of  $b_3$  in eq.(\ref{mDb1}) 
can be  parameterized in the new basis as follows:
\begin{equation}
{ \bar m_D} =m_{D0}\left ( \matrix{ 0 &  \l^{\frac{n}{2}+1} &  0\cr
\l^{\frac{m}{2}}&  0 &  0 \cr  0  & \l^{\frac{n}{2}}e^{i \rho}& 1 \cr} 
 \right ) O_R \  , 
\label{ex7}
\end{equation}
\noindent where only one phase $\rho$ remains.
The magnitude of $m_{D0}$ is determined by the relation
 $m_{D0}^2 \simeq m_0 M_3$, where $m_0\simeq \sqrt{\Delta m_{\rm atm}^2}/2$.

We examine the lepton number asymmetry in the minimal SUSY model with
the right-handed neutrinos. In the limit $M_1 \ll M_2,\ M_3$, 
the lepton number asymmetry $\e_1$ (CP asymmetry) 
for the lightest heavy Majorana neutrino ($N_1$) 
decays into $l^{\mp} \phi^{\pm}$ \cite{ep} is given by
\begin{equation}
\epsilon_1 =
\frac{\Gamma_1-\overline{\Gamma_1}}{\Gamma_1+\overline{\Gamma_1}}
\simeq  -\frac{3}{8\pi v_2^2} \left (
  \frac{{\rm Im}[\{({\bar m_D}^{\dagger} \bar m_D)_{12}\}^2]}
{({\bar m_D}^{\dagger} \bar m_D)_{11}}\frac{M_1}{M_2}  +
  \frac{{\rm Im}[\{({\bar m_D}^{\dagger} \bar m_D)_{13}\}^2]}
{({\bar m_D}^{\dagger} \bar m_D)_{11}}\frac{M_1}{M_3}\right )\ ,
\label{epsilon} 
\end{equation}
\noindent where $v_2=v \sin\beta$ with $v=174$GeV.
The lepton asymmetry $Y_L$ is related to the CP asymmetry through the relation
\begin{equation} 
Y_L=\frac{n_L-n_{\bar{L}}}{s}=\k \frac{\epsilon_1}{g_{\ast}}\ ,
\label{YL}
\end{equation}
where $s$ denotes the entropy density, 
$g_{\ast}$ is the effective number of relativistic degrees of freedom
contributing to the entropy and $\kappa$ is 
the so-called dilution factor which
accounts for the washout processes (inverse decay and lepton number violating
scattering).  In the MSSM with right-handed neutrinos, 
one gets $g_{\ast}=232.5$.

The produced lepton asymmetry $Y_L$ is converted into a net baryon asymmetry
$Y_B$ through the $(B+L)$-violating sphaleron processes.  One  finds the
relation \cite{harvey}
\begin{equation} 
Y_B=\xi \ Y_{B-L}=\frac{\xi}{\xi-1} \ Y_L \  , \qquad
\xi= \frac{8\ N_f + 4\ N_H}{22\ N_f + 13\ N_H} \ ,
\label{YB}
\end{equation}
where $N_f$ and $N_H$ are the number of fermion families and Higgs
doublets, respectively. Taking into account  $N_f=3$ and $N_H=2$ in the MSSM, 
we get 
\begin{equation}  
Y_B = -\frac{8}{15}\,Y_L \ .
\end{equation}

On the other hand, the low energy CP violation, which is a measurable 
quantity in the long baseline neutrino oscillations \cite{CP}, 
is given by the Jarlskog determinant $J_{CP}$ \cite{J}, which is calculated by
\begin{equation}  
   det [{  M_\ell M_\ell^\dagger,  M_\nu M_\nu^\dagger}]
= -2 i   J_{CP}  (m_\tau^2-m_\mu^2) (m_\mu^2-m_e^2)(m_e^2-m_\tau^2)
 (m_3^2-m_2^2) (m_2^2-m_1^2)(m_1^2-m_3^2)  ,
\end{equation}
\noindent where $M_\ell$ is the diagonal charged lepton mass matrix, 
and $m_1$, $m_2$, $m_3$ are neutrino masses.

 Since  the CP violating phase is only $\rho$, we can find a link
between the leptogenesis ($\epsilon_1$) and the low energy CP violation 
($J_{CP}$)  in our textures of the Dirac neutrinos.
By using the Dirac neutrino mass matrix in eq.(\ref{ex7}), we get
\begin{eqnarray}  
&&\epsilon_1 \simeq -\frac{3 m_{D0}^2}{8\pi v_2^2} \ \l^m  \ \sin 2\rho
\simeq -8.8 \times 10^{-17}\ M_1 \ \sin 2\rho\ \ , 
\nonumber\\
&&J_{CP}\simeq   \frac{1}{64} \l^2\
  \frac {\Delta m_{\rm atm}^2}{\Delta m_{\rm sol}^2}\   \sin 2\rho \ ,
\label{res7}
\end{eqnarray}
\noindent
where $M_1$ is given in the  $\G$ unit and $\tan\beta\geq 10$ is taken.
It is remarked that  $\epsilon_1$ only depends on $M_1$ and the phase $\rho$,
 and the relative sign of $\epsilon_1$ and $J_{CP}$ is  opposite.
Taking the experimental data
${\Delta m_{\rm sol}^2}/{\Delta m_{\rm atm}^2}\simeq \l^2$ and
$\sin 2\rho\simeq 1$,
we predict  $J_{CP} \simeq 0.01$, which is rather large and then is favored
 for the future experimental measurement.

The five cases  of the Dirac neutrino mass matrix ($a_1,a_3,b_1,c_1,c_3$) in
eqs.(\ref{mDa1}), (\ref{mDb1}) and (\ref{mDc1})  lead to same results
in eq.(\ref{res7}). In other six cases  of the Dirac neutrino mass matrix 
($a_2, a_4, b_2, b_4, c_2, c_4$), the CP violating phases are removed 
because of only three non-zero entries. Then, we get $\e_1=0$, but the same 
result in eq.(\ref{res7}) for $J_{CP}$.

If we use the modified Dirac neutrino mass matrices
in eqs.(\ref{mDa1x}), (\ref{mDb1x}) and (\ref{mDc1x}), new CP violating phases
appear.  However, the contribution to $\e_1$ is a next-leading one
as far as $x\gg 1,\ y\gg 1, z\gg 1$.

 In order to calculate the baryon asymmetry, 
we need the dilution factor involves the integration of the 
full set of Boltzmann equations \cite{pu}. 
A simple approximated solution which has been
frequently used is given by \cite{kolb}
\begin{eqnarray}
\kappa=0.3\left (\frac{10^{-3}\eV}{\tilde m_1}\right ) 
\left(\ln \frac{\tilde m_1}{10^{-3}\eV}\right )^{-0.6}\ , 
\qquad (10^{-2}\eV\leq \tilde m_1\leq 10^3 \eV ) \ ,
\end{eqnarray}
where
\begin{eqnarray}
\tilde m_1=\frac{({\bar m_D}^{\dagger} \bar m_D)_{11}}{M_1} \ .
\end{eqnarray}
By using this approximate dilution factor and  eqs.(\ref{YL}) and (\ref{YB}),
 we can  estimate  $Y_B$ in our textures as follows:
\begin{equation}
Y_B \simeq  -2.3\times 10^{-3} \ \epsilon_1 \ \kappa \ .
\label{yb}
\end{equation}
It is noticed that
 $Y_B$ and $J_{CP}$ are same sign since $\e_1$ has minus sign.
 
The  WMAP has given the new result \cite{WMAP} 
\begin{equation}
\eta_B = 6.5 ^{+0.4}_{-0.3} \times 10^{-10} \ ( 1\ \sigma) \ ,
\end{equation}
\noindent which  leads to 
\begin{equation}
Y_B \simeq  \frac{1}{7}\eta_B \ .
\end{equation}
In our textures,  we have 
$({\bar m_D}^{\dagger} \bar m_D)_{11}=m_{D0}^2\l^m$, which gives
 $\tilde m_1=\frac{1}{2}\sqrt{\Delta m_{\rm atm}^2}\simeq 0.022$.
Then we get the dilution factor $\k\simeq 7\times 10^{-3}$.
Putting the observed value into  eq.(\ref{yb}),
we get 
\begin{equation}
M_1\sin 2\rho \simeq 6\times 10^{10}\G \ .
\label{result}
\end{equation}
This result means that $M_1$ is should be larger than  $6\times 10^{10}\G$
in order to explain the baryon number in the universe.
This value is consistent with previous works \cite{lept1,lept2,lept3}. 

  It is important to present the discussion from the standpoint of the GUT,
 which is given after eq.(\ref{mDb1x}).
  Taking $n=8$ and $m=6$ in the $b_3$ case of eq.(\ref{mDb1x}) 
 as in the previous discussion, 
 one obtains $M_3\sim 10^{15} \G$ and $M_1\sim 10^{8} \G$ 
taking account of $\Delta m^2_{\rm atm}\simeq 2\times 10^{-3}\eV^2$.
This result does not satisfy the condition of  eq.(\ref{result}).
However, the simple $SO(10)$ fermion mass relation  may be consistent
 with the leptogenesis  in the case of the  more complicated texture of 
$M_R$, which leads to the two-zero texture $A_2$, as seen in the work of 
\cite{Buch}.    Details  are presented  in the preparing paper including
 the degenerate case of $M_R$ in the simple $SO(10)$ approach \cite{Bando}.

We add the discussion of  another important problem.
In the framework of supersymmetric thermal leptogenesis,
there is cosmological gravitino problems. The gravitino with
 a few TeV mass does not favor $M_1 \geq  10^{10}$ GeV
\cite{Kawasaki}, because  $M_1$ should be lower than the maximum reheating 
temperature of the universe after inflation. 
In order to keep the thermal leptogenesis in the SUSY model, we may consider
 the  gravitino with $O(100){\rm TeV}$ mass, which is derived
 from the anomaly mediated  SUSY breaking mechanism \cite{Anomaly}. 


Summary is given as follows.
We have discussed the textures with the seesaw enhancement.
These textures are important in the standpoint of the quark-lepton
unification, in which  quark masses are hierarchical and  quark mixings are 
very small.
It is very difficult to get  general conditions for the seesaw enhancement of
the bi-large mixing, however, 
the two-zero texture of the left-handed  neutrino mass matrix $M_\nu$
are  helpful  to study the seesaw enhancement of  the bi-large mixing.
Once  the basis of the right-handed Majorana neutrino mass matrix is fixed,
one can find  some sets of $m_D$ and $M_R$, which have
  hierarchical masses without large mixings, to give the two-zero textures
$A_1$ and $A_2$ without fine tuning among  parameters of these  matrices.
These sets present  the seesaw enhancement of the bi-large mixing,
because there is no large mixings in  $m_D$ and $M_R$,
but bi-large mixing is  realized via the seesaw mechanism.
We present twelve  sets of  $m_D$ and $M_R$ for the seesaw enhancement.
Then, the decay rate of  $\mu\rightarrow e+\gamma$
is  enough suppressed due to zeros in the Dirac neutrino mass matrix.
Six sets lead to  the lepton asymmetry, which depends on only $M_1$ and 
the phase $\rho$.  Putting the observed value of baryon number 
in the universe,  $M_1\simeq 6\times 10^{10}\G$ is obtained.
It is remarked that 
$J_{CP}$ is the same sign as the $Y_B$, and its  magnitude is
predicted to be $\simeq 0.01$. 
Other six ones  provide the real Dirac neutrino  mass matrices, 
which give no CP asymmetry. 
Study of modified right-handed Majorana neutrino mass matrices is
important for realistic model buildings based on the quark-lepton unification.

\vskip 1  cm
 
 This research was supported in part by the National Science Foundation 
in the Grant-in-Aid for Science Research, Ministry of Education, 
Science and Culture, Japan(No.12047220).

\newpage

\end{document}